\documentclass[11pt]{article}
\usepackage{iap2000,epsfig}

\def\Journal#1#2#3#4{{#1} {\bf #2}, #3 (#4)}

\def\be{\begin{equation}}
\def\ee{\end{equation}}
\def\bea{\begin{eqnarray}}
\def\eea{\end{eqnarray}}

\begin{document}
\vspace*{4cm}
\title{OPTICAL FOLLOW-UP OF THE REFLEX SURVEY: A NEW PERSPECTIVE IN 
CLUSTER STUDIES}

\author{L.T. BARONE$^{1,2}$, E. MOLINARI$^2$, H. B\"OHRINGER$^3$, 
G. CHINCARINI$^2$, C.A. COLLINS$^4$, L. GUZZO$^2$, P.D. LYNAM$^3$,
D.M. NEUMANN$^5$, T. REIPRICH$^3$, S. SCHINDLER$^4$, P. SCHUECKER$^3$}
\vskip0.8cm
\address{$^1$ Universit\`a degli Studi di Milano, Dipartimento di Fisica,
Via Celoria 16, 20133 Milano, Italy\\
$^2$ Osservatorio Astronomico di Brera, Via E. Bianchi 46, 23807 
Merate (LC), Italy\\
$^3$ Max-Planck-Institut f\"ur extraterrestrische Physik, 
Giessenbachstr. 1, 85740 Garching, Germany\\
$^4$ Liverpool John Moores University, Birkenhead, L41 1LD, U.K.\\
$^5$  CEA/Saclay, L'Orme des Merisiers,
 91191 Gif-sur-Yvette, France}

\maketitle\abstracts{After the completion of the catalogue built with ESO key 
program (REFLEX), we are beginning to explore the multiband optical 
characteristics of a 
subsample of those clusters selected in a statistically independent way. We 
have already observed in B, V, R, 9 of the  about 50 clusters of the 
subsample, with
$0.15 \le z \le 0.18$. 
More time has been allocated by ESO for our 
project in March 2001. We will therefore be able to compare X-ray and optical
morphologies and luminosities and, thanks to the large field of view of the
WFI instrument, assess the luminosity function and the color segregation in the
cluster on a robust statistical basis. In this poster we present the project,
the technical problems we are facing in the reduction phase,
and show the first preliminary results, which seem very encouraging.
}

\section{Introduction}
The study of individual cluster of galaxies can be considered as the repetition
of the same experiment, conducted under varying sets of initial conditions.
Unfortunately, the observer only has a comparatively poor knowledge of the
'exact' initial conditions in each cluster/experiment. As the intrinsic and
enviromental conditions cannot be chosen, a statistical approach is the only 
effective one in a proper attempt to extract the rules of the experiment, as it
is very frequently the case in astronomy.
\par
The problem of robust statistical sample selection of galaxy clusters has 
always been a concern, and only with X-ray astronomy --- a tool which is more
objective than optical identification --- at astronomers' disposal can such 
a selection be reliably performed. Furthermore, the X-ray luminosity offers 
an indirect but reliable estimate of the gravitational mass of the whole 
cluster as shown by the tight relation of the two cluster properties 
in recent work by Reiprich \& B\"ohringer\cite{rei}.
\par
The cluster characteristics observed in the optical band are the goal of the 
present project and will be analyzed and catalogued as a function of the main
X-ray properties.

\subsection{The REFLEX sample}
\begin{figure}[t]
\begin{center}
\psfig{figure=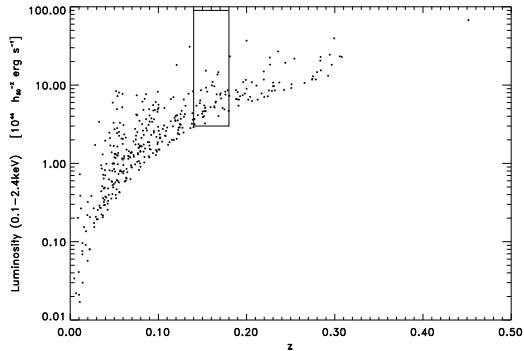,height=5cm}
\caption{Optical subsample of the REFLEX cluster survey. The rectangle 
indicates the clusters belonging to the subsample objective of our study, with
redshifts $0.15 \le z \le 0.18$.
\label{fig:lxz1}}
\end{center}
\end{figure}

A unique opportunity in this sense is provided by the recent completion of the
ESO key programme (REFLEX) aimed at the redshift measurement of all the 
candidate clusters selected in the ROSAT All Sky Survey (RASS) in the
Southern Hemisphere. This programme
(B\"ohringer {\em et al.} \cite{boh1}, Guzzo {\em et al.} \cite{guz}, 
De Grandi {\em et al.} \cite{deg}) produced a well-controlled sample of 452 
galaxy clusters down to a flux limit of $3\times10^{-12}$ erg sec$^{-1}$ 
cm$^{-2}$. Its X-ray luminosity {\it vs.} redshift distribution is reported in 
Fig. \ref{fig:lxz1}. 

\begin{figure}[b]
\begin{center}
\psfig{figure=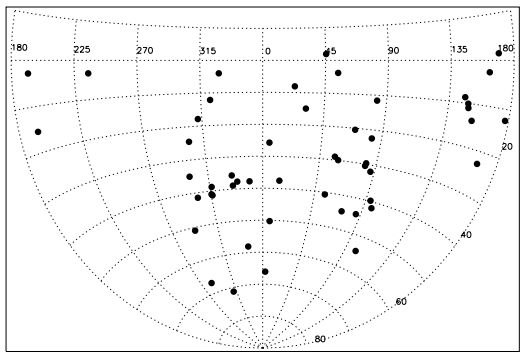,height=5cm}
\caption{Distribution on the Southern Hemisphere of the REFLEX subsample.
\label{fig:sky}}
\end{center}
\end{figure}

\subsection{The subsample}
\begin{figure}[t]
\begin{center}
\psfig{figure=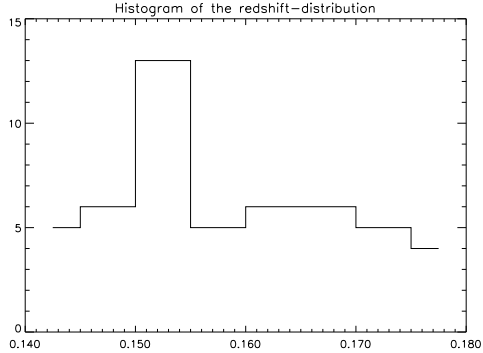,height=5cm}
\caption{Redshift distribution of the subsample.
\label{fig:lxz2}}
\end{center}
\end{figure}

A REFLEX subsample of about 50 clusters (Fig. \ref{fig:sky}) with 
redshift around 0.15 (Fig. \ref{fig:lxz2}) has thus been selected 
to begin an optical follow-up
campaign in the three bands B, V, R. Nine out of these 50 clusters have 
already been observed in two different observation runs in the 
years 1999 and 2000 at the 2.2-m ESO Telescope in La Silla (Chile).
Three more nights have been allocated for this project in March 2001.

\begin{figure}[b]
\begin{center}
\psfig{figure=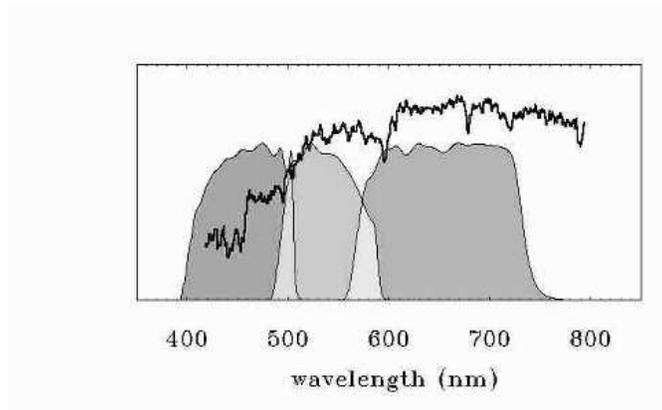,height=5.5cm}
\caption{A typical galaxy spectrum redshifted at $z=0.15$: superposed are
the 3 bands we used. The galaxy is NGC 4472, from Kennicutt $^6$.
\label{fig:filts}}
\end{center}
\end{figure}

The instrument used for this study is the WFI (Wide Field Imager), which is
a mosaic made of 8 CCDs to cover a total field of about 30'$\times$30' 
(about 8000 $\times$ 8000 pxl). At redshift 0.15, this corresponds to a
field of $5 \times 5$ Mpc$^2$ for each telescope pointing. This field
is therefore ideal to best cover all the typical virial radii of clusters
($\sim 3 h_{50}^{-1}$ Mpc) and to analyze cluster properties on scales 
far from the centre. With the exposure times used, the limit magnitude in V is
about 24. Figure \ref{fig:filts} shows, for comparison, what we expect that
a typical galaxy spectrum looks like at a redshift of 0.15. The bands
we observed fall exactly in the right range to capture the 4000-break.

\section{Our analysis}
For the observation,
we used a dithering pattern involving 6 exposures per cluster per colour
which allows in the data reduction phase a reconstruction of the whole field
of view with a fairly unifrom exposure ($\ge 80\%$ of the exposure time).
The end product shall be deep integrations of the entire virial regions
of the clusters. For the calibration of the camera, we conduct observations 
of standard stars for each of the chips in each colour and monitor the 
photometric conditions with further standard star observations during the
night. 
\par 
For the reduction, the IRAF package has been used. In particular,
the tasks {\it esowfi} and {\it mscred} are specifically designed to
face mosaic reduction of ESO WFI data.
\par
The data reduction process (still {\it in fieri}) has proved particularly
demanding.

\subsection{Astrometric correction}
At first, the problem of astrometric correction has to be faced.
The tasks used by IRAF make use of Valdes' (1998) correction. 
In order to do this, the procedure we applied is the following. Given one
of the images either of the 6 dithering sequence or of the standard star
fields, by use of {\it Skycat}, it was compared with an external star catalogue
(USNO). The comparison involves two steps: 
\begin{enumerate}
%\begin{enumerate}
\item an initial rough spatial shift on the plane, due to the fact
that the center of the CCD (and of the telescope) and of Valdes' correction
do not coincide and due to the telescope pointing imprecision; 
\item a more precise matching, which is obtained by three corrections
applied to the image: tangent point shift, fractional scale change and 
axis rotation.
%\end{enumerate}
\end{enumerate}

This latter phase is obtained by IRAF by comparing the intensity peaks of the 
image with the given catalogue. The choice of the points to exclude from the
fit is obtained by manual correction to avoid spurious ``detections'' and
minimize the rms.
\par
Subsequently, another catalogue is obtained with {\it SEXTRACTOR}
(Bertin \& Arnouts \cite{bea}) from 
the corrected image,
which is in turn used for the other images of the sequence (of dithering images
or of star fields) for the astrometric correction.
Clearly, it is the correspondence with {\it this} internal
catalogue to be important, 
since, once the dithered images are corrected, when they will have
to be stacked, it is the {\it internal} precision that will affect the 
goodness of the final (cluster) image. This second process is analogous to
the aforementioned one, except for the use of the other catalogue.
\par
\begin{figure}
\begin{center}
\psfig{figure=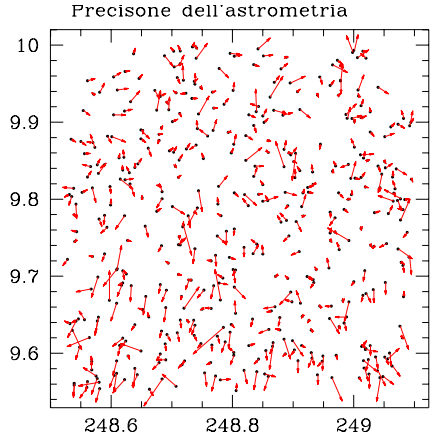,height=6cm}
\caption{The astrometric correction we performed on one of our images
with respect to the USNO catalogue (whose internal error is about 0.4 arcsec)
freed from systemic errors. On the abscissa the alpha coordinate of the image
is plotted; on the ordinate the declination is plotted. 
The arrows are enhanced 100 times.
\label{fig:sfreccia}}
\end{center}
\end{figure}

In Fig. \ref{fig:sfreccia} we show, as an example, the corrections applied 
astrometrically on one of our star field images with respect to the {\it 
external} catalogue.
The final error we obtain in the comparison with the
{\it internal} catalogue is about $\pm0.5$ arcsec, well
within an acceptable range, considering that the error in the USNO is 
about 0.4 arcsec.

\subsection{Photometric correction}

The procedure used for the subtraction of the flat field, should guarantee
that the chip-to-chip variations are taken care of. In order for us to verify
that this is actually the case, we developed a process that allows us to
check that these originally non-negligible variations are effectively reduced.
\par
\begin{figure}[t]
\begin{center}
\psfig{figure=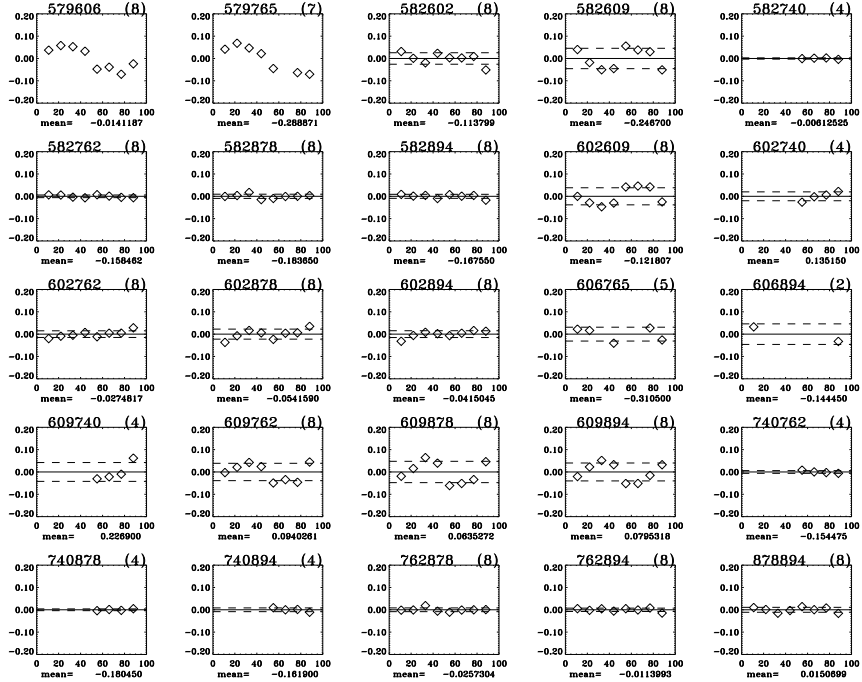,height=11.7cm,angle=+90}
\caption{This figure shows the comparison between different couples of 
catalogues coming from different exposures of the same (stellar) fields. 
Once the average displacement value from 0 has been calculated
(small caption beneath each plot), for each couple of catalogues for the
objects present in both catalogues on the {\it n}th chip (here only the
$nn$ values are shown, on the abscissa of each plot), such value has been
added to each of the value $nn$. In those cases when the variations were 
$\le 00.5$ mag, 1 $\sigma$ dashed lines have been traced 
and such values have been used to calculate the reciprocal 
efficiency of the 2 CCDs for the $nm$ cases, shown in Fig. \ref{fig:nm}.
\label{fig:vari}}
\end{center}
\end{figure}

In Fig. \ref{fig:vari} we represent the 2-by-2 comparisons between different
catalogues obtained from images of the same stellar field, for which a
comparison of the objects on the 
$n$th CCD with the same $n$th CCD in another image is 
available ($n=1,2,...,8$). These values have been averaged out and normalized
to 0 (the small number on the bottom of each plot indicates the averaged 
number that has been added to the values for this normalization). Accepted 
were only those $nn$ variations $\le0.05$ mag; $1 \sigma$ dashed lines
have then been traced.

\begin{figure}[hb]
\begin{center}
\psfig{figure=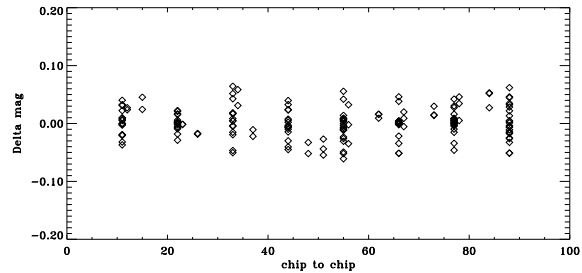,height=7.8cm,angle=+90}
\caption{In this figure all the $nm$ couples are shown, for which 
``accepted'' $nn$ values are available (Fig. \ref{fig:vari})
\label{fig:nm}}
\end{center}
\end{figure}

For those couples of catalogues for which the $nn$ values were acceptable,
the values of the calibration (excluding the normalization) for the
$nm$ CCDs (with $n \neq m$) have been calculated, as shown in Fig. 
\ref{fig:nm}.

\begin{figure}
\begin{center}
\psfig{figure=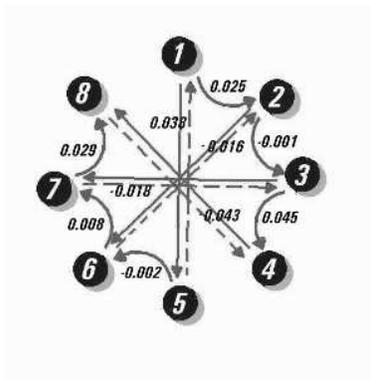,height=5cm,angle=0}
\caption{Resulting coefficients for the chip-to-chip calibrations.
Only one star field has been used so far to produce this numbers.
Calibrations still need refinment, but they already seem to show that the flat
field subtraction is sufficient to take care of the chip-to-chip variations.
\label{fig:figa}}
\end{center}
\end{figure}

Finally, on Fig. \ref{fig:figa} we sketched the numbers 
which need to be used to move from one chip to the other.
Only one star field image sequence has been used to obtain those numbers
and only one filter. Numbers still need refinment, but the results shown 
indicate that the flat field subtraction seems to be sufficient to correct
for the chip-to-chip variations.

\section{Preliminary results}

\begin{figure}[b]
%\begin{center}
\hbox to \textwidth
{
\parbox{7cm}{
\psfig{figure=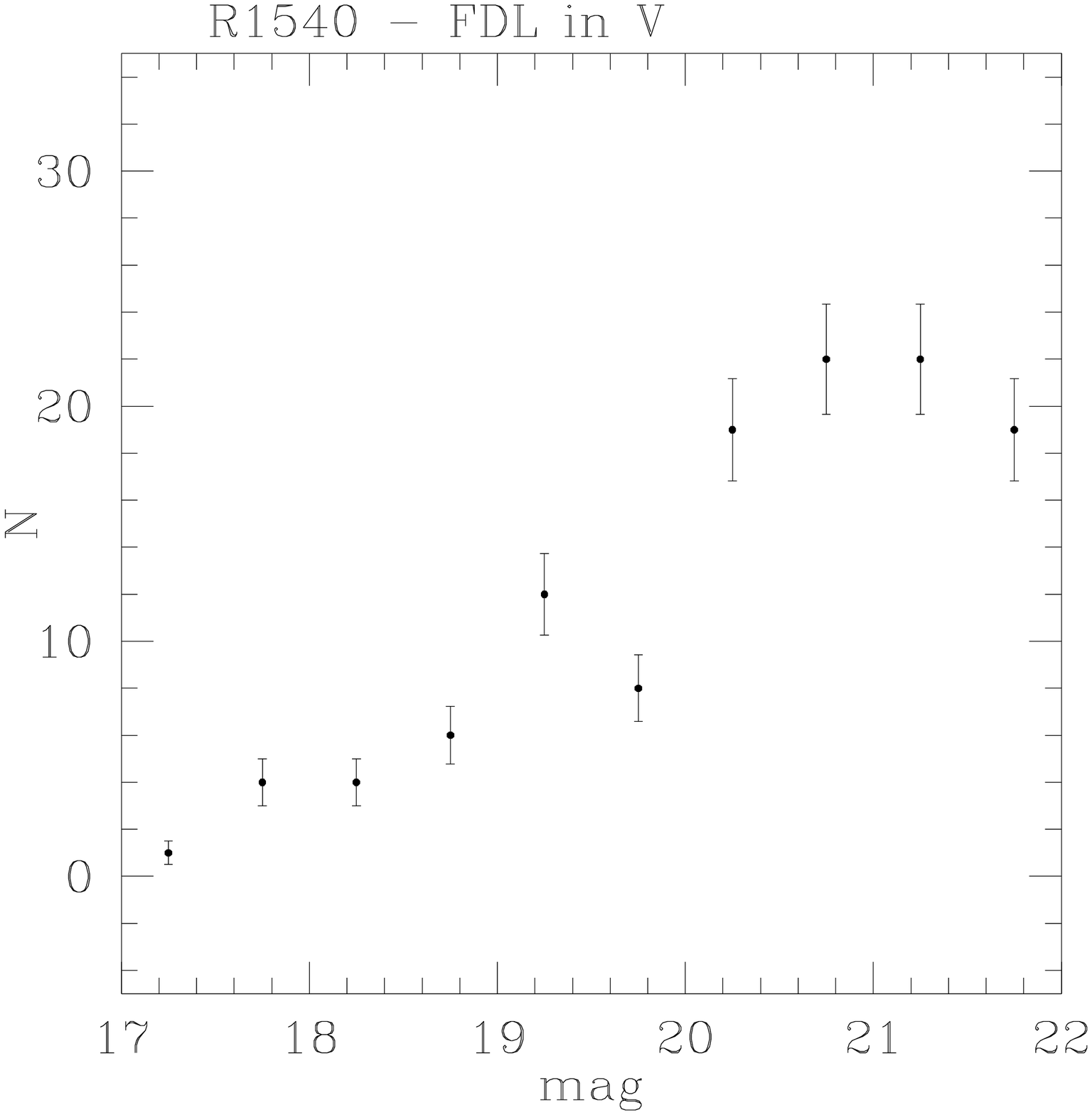,height=5cm}
\caption{Number counts of the core of cluster R1540. Errors are poissonian.
Bins are of 0.5 mag. Time of exposure is of the order of 7 min. For mag $\ge
20$ clearly we begin missing something.}
}
\label{fig:LFV}
\hfill
\parbox{7cm}{
\psfig{figure=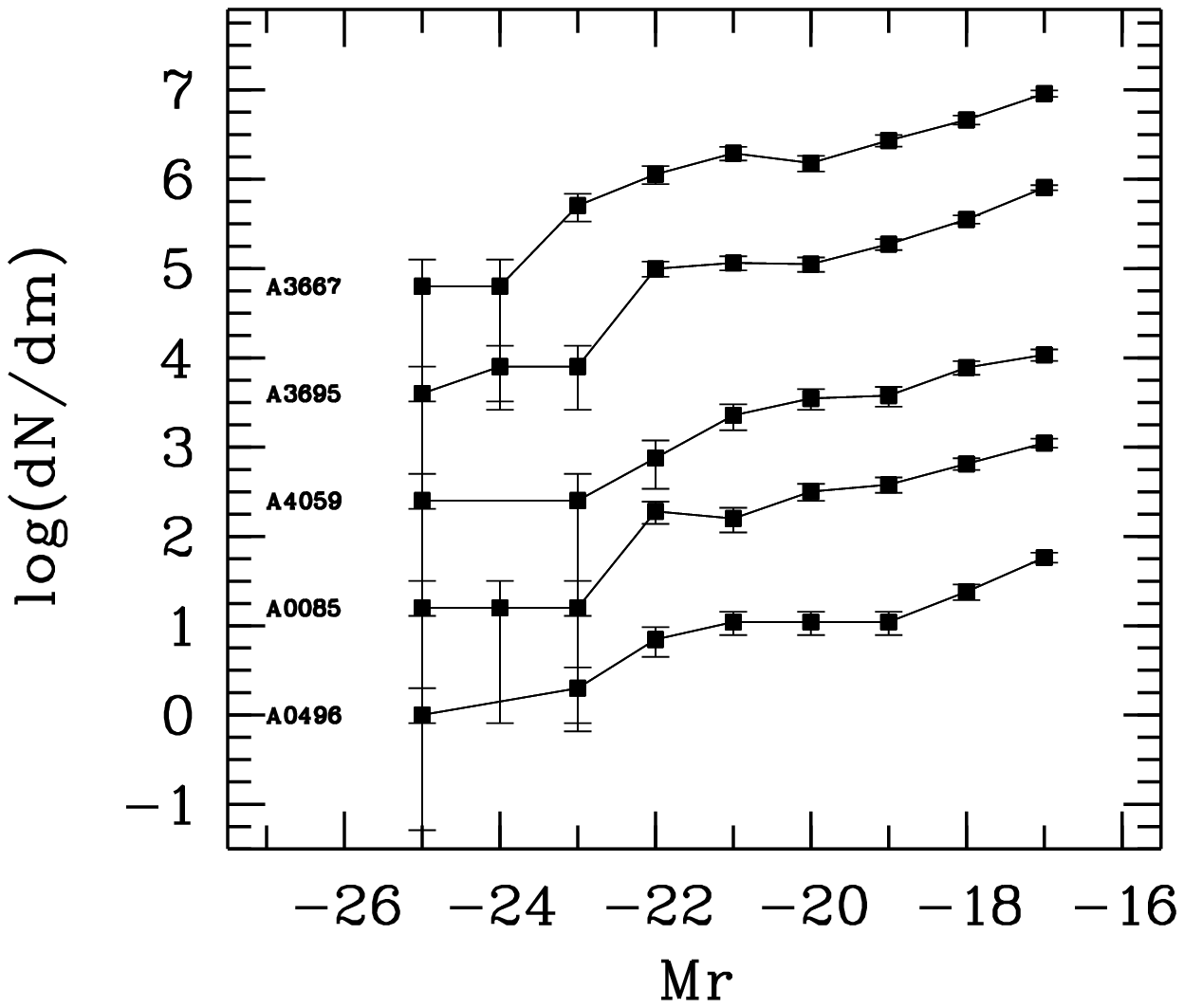,height=5cm}
\caption{This image comes form Molinari $^5$ and shows the 
bimodality between giant and dwarf ellipticals in a sample of 5 low
redshift clusters.}
}
\label{fig:emi}
}
%\end{center}
\end{figure}

A preliminary Luminosity Function in V of the area of the core of cluster 
R1540, performed on only 1 of the 6 dithered images of the cluster (1 chip)
is shown on Fig. \ref{fig:LFV}.
Even if the image is only a number count of the objects based on a 
simple {\it SEXTRACTOR} result (no backgound subtracted), it still shows an
interesting trend. For comparison, we show on the right panel of Fig.
\ref{fig:emi} a figure coming from Molinari \cite{emi}. In the
plot of the 5 low redshift clusters the bimodality separating 
the giant from the dwarf ellipticals is evident.

\begin{figure}[h]
\begin{center}
\psfig{figure=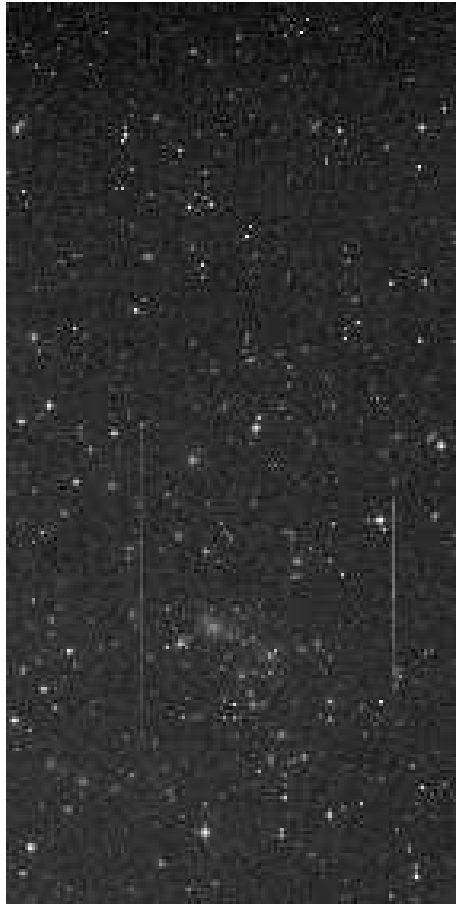,height=12cm}
\end{center}
\caption{Background of the poster. The core of the cluster R1540. The
three BVR filters have been used up to produce the image. This comes only
from 1 chip and a single exposure.
\label{fig:back}}
%\end{center}
\end{figure}

The core of cluster R1540 is the one shown in the background image of the 
poster presented (Fig. \ref{fig:back}).

\section*{Acknowledgments}
We would like to thank S. Benetti and A. Zacchei (TNG staff) for help in
producing the poster background image (Fig. \ref{fig:back}) 
and L. Rizzi for his 
constant help and precious tips and suggestions in the data reduction 
process. We are also very grateful to A. Moretti for valuable help 
in the clusters observations.

\end{document}